\documentstyle[aps,epsf,aps10]{revtex}
\begin{document} 

\title{Radiative Transitions in Heavy Mesons\\
in a Relativistic Quark Model}
\author{J. L. Goity}
\address{
Department of Physics, Hampton University, Hampton, VA 23668, USA \\
and \\
Thomas Jefferson National Accelerator Facility  \\
12000 Jefferson Avenue, Newport News, VA 23606, USA.}
\author{W. Roberts}
\address{National Science Foundation, 4201 Wilson Boulevard, Arlington, 
VA 22230
\footnote{on leave from Department of Physics, Old Dominion University, 
Norfolk, VA
23529, USA, and
 Thomas Jefferson National Accelerator Facility,
12000 Jefferson Avenue, Newport News, VA 23606, USA.}.}
\date{\today}
\maketitle 
\begin{abstract}
The radiative decays of $D^*$, $B^*$, and other excited heavy mesons are 
analyzed in 
a relativistic quark 
model for the light degrees of freedom and in the limit of heavy quark 
spin-flavor symmetry.
The   analysis of
strong decays carried out in the corresponding chiral quark model is used
to calculate the strong decays and determine the branching ratios of
the radiative $D^*$ decays. Consistency with the observed branching ratios
requires  the inclusion of the heavy quark component of the
electromagnetic current and the introduction of an anomalous magnetic moment
for  the light quark. It is observed that not only $D$, but also  $B$ meson
transitions within a heavy quark spin multiplet are affected by the presence
of the heavy quark current. \vspace{5mm}

\flushright{JLAB-THY-00-45}
\end{abstract}
\thispagestyle{empty}
\pacs{\tt$\backslash$\string pacs\{13.25.-k, 12.39.-Ki, 12.39.Fe, 
12.39.Hg \}} 
\setcounter{page}{1}

\section{Introduction}

In recent articles, we examined the spectra \cite{vanorden} and strong decay 
widths
\cite{GR1} of a number of heavy-light mesons 
in a relativistic chiral-quark model. The strong decays were  assumed to take 
place
through pion or kaon  emission from the brown muck of the heavy
meson, with the heavy quark being essentially a spectator in the decay. We 
found
that relativistic effects in the decays were  large, as many results we
obtained were quite different from those obtained in analogous non-relativistic
calculations \cite{GR2}.

It is well known that the meson emission decays of the ground-state vector 
mesons are
suppressed or forbidden by phase space. 
For the $D^*$ mesons, the measured radiative partial
widths are of the same order of magnitude as their partial widths for pion
 emission as a
result of this suppression. The branching ratios reported are \cite{pdg}
\begin{eqnarray}
&&{\rm BR}(D^{* 0}\to D^0\pi^0)=61.9\pm 2.9\%,\nonumber\\
&&{\rm BR}(D^{* 0}\to D^0 \gamma)=38.1\pm 2.9\%,\nonumber\\
&&{\rm BR}(D^{* +}\to D^+ \pi^0)=30.6\pm 2.5\%,\nonumber\\
&&{\rm BR}(D^{* +}\to D^0 \pi^+)=68.3\pm 1.4\%,\nonumber\\
&&{\rm BR}(D^{* +}\to D^+ \gamma)=1.1^{+2.1}_{-0.7}\%,\nonumber\\
&&{\rm BR}(D_s^{* }\to D_s \pi^0)=5.8\pm 2.5\%,\nonumber\\
&&{\rm BR}(D_s^{* }\to D_s \gamma)=94.2\pm 2.5\%.\nonumber
\end{eqnarray}
In contrast with this, the total widths of these 
states are not experimentally known, with only  upper limits reported 
\cite{pdg}:
\begin{eqnarray}
&&\Gamma_{D^{*\pm}}<131 {\rm keV},\nonumber\\
&&\Gamma_{D^{* 0}}<2 {\rm MeV}.\nonumber
\end{eqnarray}
The use of isospin symmetry arguments for the strong decay amplitudes, 
together 
with the reported
branching ratios of the $D^{*0}$, suggest the rough upper limit 
$\Gamma_{D^{* 0}}<75$ keV.
Finally, no information is currently available
on other radiative transitions, except for the transition $D_{s1}(1^+)\to 
D_s^*\gamma$ quoted as `seen' by the Particle Data Group
\cite{pdg}.

In addition to the states mentioned above, the electromagnetic decays of a 
number of the 
lower lying states, including orbitally excited states, are expected to be
significant or even dominant. 
One example are  the $B^*$ mesons,  where the $B^*-B$ mass 
differences are of the order of 50 MeV, leaving the radiative mode as the only 
possible one. 
Note that no experimental information on the $B^*$ radiative decays is 
available. 
For some of the orbitally excited $D_s$ and 
$B_s$ states, $K$ meson emission is forbidden or suppressed by phase space, 
leaving the electromagnetic decays and  isospin violating and/or OZI
violating pion emisssion, as the only possible  decay modes. Examples of such
states are the $D^*_s$ and the  $D_s \, (0^+,1^+)$ and 
$B_s \, (0^+,1^+)$ doublets. A calculation of the radiative  decays is thus 
crucial for
understanding these  states.

The radiative decays of heavy mesons have been examined by a number of authors, 
in a variety of
different frameworks. Kamal and Xu \cite{kamal} have treated the decays of the 
ground state vector mesons in a simple quark model. Fayyazuddin and Mobarek 
\cite{mobarek} have 
also used a simple quark model to study these decays, but have also looked at 
the radiative 
decays of the  $(1^+,2^+)$ multiplet. Oda, Ishida and Morikawa \cite{oda} have 
used a covariant
oscillator model, while Avila \cite{avila} uses a covariant model to describe 
the mesons, and both
sets of authors deal only with the $M_1$ decays of the ground state vector 
mesons. Cho and Georgi
\cite{cho}, Cheng {\it et al.} \cite{cheng}, and Amundson {\it et al.} 
\cite{amundson} use the heavy hadron chiral symmetry to
treat these decays, while K\"orner {\it et al.} \cite{korner} use the heavy 
quark effective theory in a hybrid
approach, supplemented with quark model input, to treat the decays of the 
$(1^+,2^+)$ multiplet.
L\"ahde, Nyf\"alt and Riska \cite{riska} treat the decays within the formalism 
of the
Blankenbecler-Sugar equation. Prior to the work mentioned above, Pham 
\cite{pham}, 
Rosner \cite{rosner}, Miller and Singer \cite{miller}, 
and Eichten {\it et al.} \cite{eichten} have all examined these decays in 
different scenarios.
 
\section{Model}
In this work  we study those radiative transitions that have been  or could eventually be
experimentally observed. Our model is the same relativistic chiral quark model
used in $\cite{GR1}$
where the emission of light pseudoscalars was studied. 
In that model, the only input beyond the potential and the constituent quark masses is  the 
axial vector coupling of the constituent quark, $g_A^q$. 
For the radiative transitions one has to consider the electromagnetic
current. In the large heavy-quark mass limit the dominant contributions to
radiative transitions come from the light-quark component of the current,
however, as explained later, the $D^*$ radiative decays are very much
affected by the contributions of the heavy-quark current. The heavy-quark 
component of the current couples to the photon through its spin piece which is
suppressed by one power of $1/m_Q$, leading to a suppression of the heavy quark
contributions to the radiative amplitudes  by at least one power of  $1/m_Q$.
The electromagnetic current in the model is \begin{equation} J^\mu=
J^\mu_q+J^\mu_Q=  e\;\bar{q}\; \hat{\bf {\cal Q}}\;
(\gamma^\mu+\frac{1}{\Lambda}\sigma^{\mu\nu}
\partial^{^{\!\!\!\!\leftrightarrow}}_\nu) q+e\;
\bar{Q}\;\hat{\bf {\cal Q}}\;\gamma^\mu Q, \end{equation} where we have included an 
anomalous magnetic moment for the light quark. Here  $\hat{\bf {\cal Q}}$ is the
electric  charge operator, and  $\Lambda= 2\tilde{m}_q/\kappa^q$ with
$\tilde{m}_q$ the constituent light quark mass. In this work we use $
\tilde{m}_u= \tilde{m}_d=253$ MeV, $ \tilde{m}_s=450$ MeV, and the heavy
quark masses are taken to be $m_c=1.53$ GeV and $m_b=4.87$ GeV.

It is instructive to examine first the contributions from the heavy quark current to the radiative amplitudes
within the framework of the heavy quark effective theory (HQET). Consider the transition 
between $D$ mesons $D_a\to
D_b\gamma$. Let us define
\begin{equation}
M_a=m_c+\Lambda_a,\,\,\,\, M_b=m_c+\Lambda_b,
\end{equation}
and let the velocities of the hadrons be $v_a$ and $v_b$,
respectively, with the four-momentum of the photon being $K$. In the rest frame of the parent hadron, and 
to order $1/m_c^2$,
\begin{eqnarray}\label{hqeta}
v_a\cdot K&=&\left(\Lambda_a-\Lambda_b\right)\left[1-\frac{\Lambda_a-\Lambda_b}{2m_c}
+\frac{\Lambda_a\left(\Lambda_a-\Lambda_b\right)}{2m_c^2} + \dots\right],\nonumber\\
v_a\cdot v_b&=&1+\frac{\left(\Lambda_a-\Lambda_b\right)^2}{2m_c^2}+\dots.
\end{eqnarray}
 For transitions within a multiplet, such as for $D^*\to D\gamma$,
the mass difference $\Lambda_a-\Lambda_b$ is of order $1/m_c$.

Assuming that the initial meson is at rest, and that the emerging photon defines the $z$-axis, we can write
\begin{equation}\label{hqetb}
v_a=(1,0,0,0), \,\,\, v_b=
\left(1+\frac{\left(\Lambda_a-\Lambda_b\right)^2}{2m_c^2},0,0,
-\frac{\left(\Lambda_a-\Lambda_b\right)}{m_c}\left(1-\frac{\Lambda_a+\
\Lambda_b}{2m_c}\right)\right). \end{equation}
With all this we can now obtain  the suppressions in $1/m_c$ of the different  
matrix elements of the heavy quark   current with respect to the light quark
one.  
Since we are interested  in the emission of  a real photon, only
those pieces of the current matrix elements that can couple to a real photon
are of interest to us. There is a generic  $1/m_c$ suppression due to the
magnetic moment of the heavy quark. For  transitions within a heavy-quark spin
multiplet this is the only suppression, while   for transitions between states
in two different multiplets the suppression is instead $1/m_c^n$, where 
$n=\max\{3,\; 1+|\ell-\ell'|\}$ if the states have the same parity, and 
$n=1+|\ell-\ell'|$ if the states have opposite parity. Here, one power of
$1/m_c$ is the one mentioned before and the rest stems  from  the overlap of
the initial state light quark wave function with the boosted one of the final
state. Table \ref{suppressions} shows the forms of the leading order (in HQET)
matrix elements  for a few selected transitions. Note that with the form of the
matrix elements used here the Isgur-Wise form factors are of zeroth order in
the $1/m_c$ expansion. Also shown in the table is the suppression with respect
to the corresponding  light quark current matrix element. Note that for the decays between any 
given pair of heavy-quark spin multiplets, all of the matrix elements of the heavy-quark current 
appear at the same order in the heavy quark expansion. 

For transverse photons, current conservation implies that terms
in the matrix elements  of the currents that are proportional to
$v_{a,b}^{\mu}$ do not contribute to the amplitude. As discussed previously,
the recoil factor $(v_a\cdot v_b-1)$ is of  order $1/m_c^2$, while terms like
$\epsilon(v_a)\cdot v_b$ and $\epsilon(v_b)\cdot v_a$ either vanish, or are of
order $1/m_c$.  Moreover, for decays within a multiplet,
the mass difference is also of order $1/m_c$, providing a further suppression for 
such decays. In the case of the
decays from the radially excited $(0^-,1^-)^{\prime}$ multiplet to the ground state 
multiplet the matrix element has a form analogous to that shown in the first
row of Table \ref{suppressions}, except for the 
extra suppression factor $1/m_c^2$ that results from the overlap of the
wave functions. In summary, for the transitions $D^*\to
D\gamma$ as well as all other transitions we consider, the contributions
of the heavy quark to the current are subleading in the heavy quark expansion,
being suppressed by one power of $1/m_c$ in the case of transitions within a
heavy-quark spin multiplet, and at least two powers of $1/m_c$  in the case of
transitions between different multiplets. Since in the case of $D$ mesons the
charge of the heavy quark is 2/3 and $m_c$ is not  very large, one has to 
keep the contributions suppressed by $1/m_c$. Indeed, if one would disregard
these contributions, one immediately  finds  an inconsistency in the radiative
branching ratios for the non-strange $D^*$-mesons. From isospin symmetry in
the pion emission amplitudes and the relation $\Gamma(D^{*+}\to
D^+\gamma)=\Gamma(D^{*0}\to D^0\gamma)/4$, that results when
the heavy-quark electromagnetic current is disregarded, along with the corresponding
neutral pion emission branching ratios, one obtains ${\rm BR}(D^{* +}\to
D^+ \gamma)/{\rm BR}(D^{* 0}\to D^0 \gamma)\simeq 0.18$,
which is much larger than the experimental ratio $0.03^{+ 0.05}_{- 0.02}$. As
we see later, the inclusion of the heavy quark contributions largely remedies
this discrepancy.  We also find that even the $B$ meson transitions within a
heavy-quark spin multiplet are noticeably affected by  the presence of the
heavy-quark current. Thus, throughout we will keep the contributions of the
heavy-quark current to the intra-multiplet transitions. 

In the following we work in the same framework as our previous paper $\cite{GR1}$. 
We write the wave function of the light valence quark as
\begin{eqnarray}
\psi_{j l m} &=&  \left(
 \begin{array}{c} i  F(r)\; \Omega_{j l m}\\
 \\
  G(r) \;\Omega_{j\tilde{l} m}
 \end{array} \right), \nonumber\\
 \tilde{l} &=& 2j-l,
 \end{eqnarray}
 where the radial wave functions are real, and the spinor harmonics are
given by
 \begin{eqnarray}
\Omega_{(j=l+\frac{1}{2})\,l\, m} &=& \left(
 \begin{array}{c} \sqrt{\frac{j+m}{2 j}}\;Y_{l\,m-\frac{1}{2}}\\
 \\
\sqrt{\frac{j-m}{2 j}}\;Y_{l\,m+\frac{1}{2}}
 \end{array} \right), \nonumber\\
\Omega_{(j=l-\frac{1}{2})\,l\, m} &=& \left(
 \begin{array}{c} \sqrt{\frac{j+1-m}{2
(j+1)}}\;Y_{l\,m-\frac{1}{2}}\\
 \\
  -\sqrt{\frac{j+m+1}{2 (j+1)}}\;Y_{l\,m+\frac{1}{2}}
 \end{array} \right).
\end{eqnarray}
We follow here the conventions of Bjorken and Drell
$\cite{BjDrell}$.

Straightforward evaluation of the matrix elements of the electromagnetic current in the rest frame of the heavy 
meson gives
\begin{eqnarray}
&<J',M',j',l'| J^0_{q}|J,M,j,l>&=
\sum_{\ell} (-i)^{\ell} Y_{\ell (M'-M)}^*(\hat{k})
<\ell, M'-M; J,M|J',M'> \nonumber\\&&\times
T^q_0(k,\ell,J,j,l,J',j',l',\Lambda),
\end{eqnarray}
\begin{eqnarray}
&&<J',M',j',l'| J^m_{q}|J,M,j,l>=
\sum_{\ell,\ell'}(-i)^{\ell}  Y_{\ell (M'-M-m)}^*(\hat{k}) <\ell', M'-M; J,M|J',M'>
\nonumber\\
&&\times
<1,m;\ell,M'-M-m|\ell', M'-M>
T^q_1(k,k^0,\ell,\ell', J,j,l,J',j',l',\Lambda),
\end{eqnarray}
and
\begin{eqnarray}
&&<J',M',j',l'| J^m_{Q}|J,M,j,l>=i \; Y^*_{1(M'-M-m)}(\hat{k}) 
 <1, m; 1, M'-M-m|1, M'-M>\nonumber\\&&\times
<1, M'-M, J,M| J', M'> 
T^Q_1( J,j,l,J',j',l'),
\end{eqnarray}
where $k=|\vec{K}|$ and $k_0=K_0$. Here, we use the angular momentum projection basis, so that
$m=\pm1, 0$. 
Using the standard notations for the $3j$, $6j$ and $9j$ symbols, we define two reduced matrix elements 
$R_0$ and $R_1$ as
\begin{eqnarray}
R_0(j,l,j',l',\ell)&=&(-1)^{j+\ell+1/2} \frac{1}{\sqrt{4 \pi}}
\sqrt{(2 \ell+1) (2 l+1) (2 l'+1) (2 j+1) (2 j'+1)}\nonumber\\&\times&
\left(
 \begin{array}{ccc} \ell & l &l'\\
  0&0&0   
 \end{array} \right)
\left\{
 \begin{array}{ccc} \ell & l & l' \\
   1/2& j'& j  
 \end{array} \right\},
\end{eqnarray}
and
\begin{eqnarray}
R_1(j,l,j',l',\ell,\ell')&=&
(-1)^{j+j'+l} \sqrt{\frac{3}{2 \pi}} \sqrt{(2 \ell+1) (2 l+1) (2 l'+1)(2 j+1) (2 j'+1)}\nonumber\\&\times&
\left(
 \begin{array}{ccc} \ell & l &l'\\
  0&0&0   
 \end{array} \right)
\left\{
 \begin{array}{ccc} 1/2 & l & j \\
   1/2& l'& j' \\
  1 & \ell & \ell'\\
 \end{array} \right\}.
\end{eqnarray}
With this, the expressions for the reduced amplitudes $T_0$ and $T_1$ are
\begin{eqnarray}
&&T^q_0(k,\ell,J,j,l,J',j',l',\Lambda)=
4 \pi e{\bf {\cal Q}}_q \sqrt{2 J+1} (-1)^{J'} 
\left\{
 \begin{array}{ccc} \ell & j & j' \\
   1/2& J'& J
 \end{array} \right\}\nonumber\\&\times&
\left\{(-1)^{j+1/2} \int r^2 j_{\ell}(k r)
\left[(-1)^{l+l'} F'(r) F(r) R_0(j,l,j',l',\ell)\right.\right.\nonumber\\
&+&\left.(-1)^{\tilde{l}+\tilde{l'}} G'(r) G(r)
R_0(j,\tilde{l},j',\tilde{l'},\ell)\right]\nonumber\\& +&
i \frac{k}{\Lambda }\sum_{\ell'}(-i)^{\ell'-\ell} (-1)^{j'+1/2}
\sqrt{(2 \ell+1) (2 \ell'+1)} 
\left(
 \begin{array}{ccc} \ell' & 1 &\ell\\
  0&0&0   
 \end{array} \right)\nonumber\\&\times&
\int r^2 j_{\ell'}(k r)
\left[F'(r) G(r) (-1)^{\tilde{l}+l'} R_1(j,\tilde{l},j',l',\ell',\ell)\right.
\nonumber\\&+&
\left.\left. G'(r) F(r) (-1)^{\tilde{l'}+l} R_1(j,l,j',\tilde{l'},\ell',\ell)\right]
\vphantom{\int}\right\},
\end{eqnarray}
\begin{eqnarray}
&&T^q_1(k,k_0,\ell,\ell', J,j,l,J',j',l',\Lambda)=
4 \pi e {\bf {\cal Q}}_q \sqrt{2 J+1} \sqrt{2 \ell'+1} (-1)^{J'}
\left\{
 \begin{array}{ccc} \ell' & j & j' \\
   1/2& J'& J
 \end{array} \right\}\nonumber\\&\times&
\left\{(-1)^{\ell'+j'+1/2}\int r^2 j_{\ell}(k r)
\left[-i F'(r) G(r) (-1)^{\tilde{l}+l'} R_1(j,\tilde{l},j',l',\ell,\ell')\right.\right.
\nonumber\\&+&
\left.i  G'(r) F(r) (-1)^{\tilde{l'}+l} R_1(j,l,j',\tilde{l'},\ell,\ell')\right]
\nonumber\\&+&
i \frac{k_0}{\Lambda} (-1)^{\ell'+j'+1/2}
\int r^2 j_{\ell}(k r)
\left[F'(r) G(r) (-1)^{\tilde{l}+l'} R_1(j,\tilde{l},j',l',\ell,\ell')\right.
\nonumber\\&+&
\left.G'(r) F(r) (-1)^{\tilde{l'}+l} R_1(j,l,j',\tilde{l'},\ell,\ell')\right]
\nonumber\\&-&
\sqrt{2} \frac{k}{\Lambda}
\sum_{L}(-i)^{L-\ell} (-1)^{j'+1/2+L+\ell}\sqrt{3 (2 \ell+1) (2 L+1)}
\left(
 \begin{array}{ccc} L & 1 &\ell\\
  0&0&0   
 \end{array} \right)
\left\{
 \begin{array}{ccc} 1 & 1 & 1 \\
   L& \ell'& \ell
 \end{array} \right\}\nonumber\\&\times&
\int r^2 j_{L}( k r)
\left[(-1)^{l+l'}F'(r) F(r) R_1(j,l,j',l',L,\ell')\right.
\nonumber\\&-&
\left.\left.(-1)^{\tilde{l}+\tilde{l'}} G'(r) G(r)
R_1(j,\tilde{l},j',\tilde{l'},L,\ell')\right]\vphantom{\int}\right\}
\end{eqnarray}
and
\begin{eqnarray}
&&T^Q_1(J,j,l,J',j',l')=i\; k\, (-1)^{J+j-1/2} 
 \delta_{l l'} \delta_{j j'} \frac{e {\bf {\cal Q}}_Q}{m_Q}\sqrt{4 \pi} \sqrt{2 J+1}
\left\{
 \begin{array}{ccc} 1 & \frac{1}{2} & \frac{1}{2} \\
   j& J'& J
 \end{array} \right\}.
\end{eqnarray}
Note that we have kept only the leading contribution to $T^Q_1$ in the $1/m_Q$ expansion.
In particular, this means that $T^Q_1$  must vanish unless the initial and final 
states belong to the same heavy-quark spin multiplet.
 The radial wave  functions $F$ and $G$ are those used in reference
\cite{GR1}, which were obtained  using the potential and parameters in
reference \cite{vanorden}. We  have explicitly checked that the  matrix
elements of the electromagnetic current given  above do satisfy the
constraints imposed by current conservation and the relations implied by heavy
quark spin-flavor symmetry.

In terms of the reduced amplitudes $T_0$ and $T_1$, the radiative decay widths 
are written as 
\begin{eqnarray}
\Gamma_\gamma(Jjl\to J'j'l')&=& \frac{2 J'+1}{2 J+1}\frac{E'}{E}\frac{k}{8 \pi^2}
 \sum_{\ell=|J-J'|}^{J+J'}\left[\vphantom{\sum_{\ell'=\ell -1}^{\ell +1}}
 -\left|\vphantom{\frac{a}{b}}T^q_0(k, \ell, J,j,l,J',j',l',\Lambda) \right|^2\right.
\nonumber\\
&+&
\left.\sum_{\ell'=\ell -1}^{\ell +1} 
\left|\vphantom{\frac{a}{b}}T^q_1(k,k_0, \ell,\ell',  J,j,l,J',j',l',\Lambda)+
\delta_{\ell 1}\delta_{\ell' 1}T^Q_1(J,j,l,J',j',l')\right|^2\right].
\end{eqnarray}

\section{Results}
 
In three tables we show the results obtained using the model. 
The uncertainties shown in the tables
are theoretical kinematic uncertainties due to the uncertainties in the masses of the
observed  states. For
states that have not yet been observed, the uncertainties in the masses 
are taken as $\pm 20$ MeV. 
In each table, we present results for three
values of $\kappa^q$. The two non-zero values are chosen to reproduce as well as 
possible  the experimentally reported branching
fractions. The value of $\kappa^q=0.55$ is obtained from a fit to these 
branching fractions, keeping $g_A^q$ constant at a value of
0.8. The value of $\kappa=0.45$ is chosen to illustrate the sensitivity to this quantity. 
We emphasize here that
the value of $\kappa^q$ is the same for all light  flavors. Since $\kappa^q$ is
determined solely by the strong interaction, 
it is an SU(3) singlet, except for  SU(3) breaking corrections
of order $(m_s-m_{u,d})/\Lambda_{{\rm QCD}}$, which we disregard here.

Table \ref{nonstrange} shows the results obtained for the non-strange heavy mesons. 
Since all these mesons
decay primarily through pion emission, most of the radiative decays are not likely to 
be measurable.
 We show only the intra-multiplet decays, but these are also unlikely to be
measured. Among these, the ones of real interest are the  decays within the
ground state multiplet. As mentioned above, if we choose $g_A^q$=0.8, and use
the reported ratios of widths for the $D^*$ mesons, the fit value is
$\kappa^q=0.55$, while $\kappa^q=0.45$ also gives a reasonable description of
the  data. The implied total widths of the $D^{*0}$ are then 74 keV and 68
keV,  respectively, for the two values of $\kappa^q$, very close to
the  estimated upper bound of about 70 keV mentioned earlier. For
the $D^{*+}$, we obtain a total pion emission width of 93 keV, and for  the
non-zero values of $\kappa^q$ the radiative branching ratio is either 1.5 \%
or 1.0 \%, both close to the experimentally measured value. The corresponding total width is 
about 95 keV. 

The heavy quark current plays an important role in the radiative decays of the $D^*$ mesons,
 providing a reduction of the radiative widths of the $D^{*+}$ and $D^*_s$.
Indeed,  there is a large cancellation between
the heavy and light quark contributions (they add up in the  $D^{*0}$). We
observe that the light quark contribution receives a suppression because the
photon momentum is not that small (140 MeV). The $D^*$ and $B^*$ decays
are found to be very  sensitive to the anomalous magnetic moment of the light quark. 
 While
in a non-relativistic approximation the anomalous magnetic moment contribution
to the width manifests itself in the factor $(1+\kappa^q)^2$,  relativistic
effects turn out to
 give  an enhancement of the anomalous magnetic moment piece by roughly
a factor of four.  For the other decay widths that are predominantly of $M_1$
type, which includes all the intra-multiplet decays, there is a similar
sensitivity to the anomalous magnetic moment.

The widths obtained for the $B^{*}$ mesons are quite small, as one
would expect from the smaller available phase space than in the $D^*$
mesons. It is evident from the fact that the ratio of the $B^+$ to the $B^0$
width is not approximately equal to  four, that the heavy quark current has a substantial
effect. If the light
quark has no anomalous magnetic moment, the $B^{*0}$ width is found to be 40
eV, and it rises to 244 eV when $\kappa^q=0.55$. The $B^{*+}$ width is a
about a factor two to three larger than the $B^{*0}$ width. 
The intra-multiplet partial widths of the excited mesons are negligible with
respect to their radiative widths for decay into the ground state mesons.
We display them only for the sake of completeness.

Table \ref{otherresults} shows the branching ratios and widths obtained in this model. 
Also shown in that table are some
representative results presented in the literature. All of the models predict similar 
branching ratios, but there is some spread in
the predicted widths, especially for the $D^{*0}$.

The results that we obtain for the strange heavy mesons $D_s$ and $B_s$ are shown in 
table \ref{strange}.
There, we list only those decays for states whose kaon emission decays are forbidden or suppressed 
by phase space. We
also show the intra-multiplet decays. Here, as in the non-strange $D$ mesons,  the results 
are very sensitive to the value of
$\kappa^q$. In addition, states like the $(0^+,1^+)$ doublet, which would be broad if 
kaon emission could
take place, are predicted to be a few tens of keV wide. Other states, like the radially excited
$(0^-,1^-)$ doublet, are of the order of 10 keV in width. The branching ratios reported for 
the $D_s^*$ would give total widths for this state of about 107 eV, 175 keV
or 341 keV, depending on if we use $\kappa^q=0$, $\kappa^q=0.45$ or
$\kappa=0.55$, respectively. These numbers imply $\Gamma(D^*_s\to {\rm
D}_s \pi^0)=6.2\pm 2.2$ eV, $10.1\pm 3.5$ eV or $19.7\pm 7.0$ eV.  We should
observe  here that the prediction of our model for the $ D^{*+}$
radiative branching ratio and the observed  strong decay branching ratio of
the  $D^*_s$  give that ${\rm BR}(D^*_s\to D_s
\pi^0)\cdot{\rm BR}(D^{*+}\to D^+ \gamma)$ is about $2\times
10^{-3}$, a result that is much larger than the proposed in reference
\cite{cho}, where this product is  estimated in a model of the isospin
violating decay $ D^*_s\to D_s \pi^0$ and found to be about
$8\times 10^{-5}$. From  the experimental branching ratios one obtains 
$6.4\pm 4.9 \times 10^{-4}$, which falls between those two numbers.  

The strong decays of many of the excited strange heavy mesons will proceed either
through the emission of one or two  pions; only the $(1^-,2^-)$,
$(0^{-'}, 1^{-'})$, and the $(1^+,2^+)$ states can decay emitting a kaon. 
The $(1^+,2^+)$ states have masses that lie very close to the threshold for kaon emission. However,
these decays are expected to be predominantly $D$-wave, so that the centrifugal suppression will lead
to very small decay widths \cite{GR1}. Similarly, in the $(0^{-'}, 1^{-'})$ multiplet, the
$0^{-\prime}$ may actually lie below the threshold for kaon production, meaning that its
electromagnetic decays could provide a significant portion of its total decay rate. 
The pionic decay widths of these states
have not been studied to the best of our knowledge. We expect them to be in
the range from a few tens to a hundred keV. Thus, several of the excited states
will  decay radiatively with an important branching fraction.

We note that heavy-quark  spin-symmetry would require that
both partners in a heavy quark spin multiplet should have the same partial width 
for transitions to other multiplets. In the table this holds only 
approximately because of phase space corrections due to
the subleading intra-multiplet mass splittings. In a few cases the effect is
dramatic  because the phase space is small to start with.

\section{Conclusions}
In summary, we have obtained results for a variety of radiative heavy meson
transitions in the relativistic quark model. To agree with observed decays
it is found that the light quark must have an anomalous magnetic moment
of about 0.5. Important corrections subleading in the expansion in the
inverse of the heavy quark mass are observed. These corrections are very
important in the $D$-meson sector for transitions within the same heavy quark
spin multiplet, while in $B$ mesons those contributions are much smaller but not
quite negligible. The decays of excited  $D_s$-mesons are particularly 
interesting as their strong decays are suppressed. The experimental study of some of the radiative and  strong
decays would clearly impact on our understanding of the structure of
heavy mesons and their excited states. \\\\

This work was
supported by the National Science Foundation through grants
\# PHY-9733343 (JLG), \# PHY-9457892 and \# PHY 9820458 (WR),
 by the Department of Energy through contracts DE-AC05-84ER40150 (JLG and WR) 
and DE-FG05-94ER40832  (WR), and by  sabbatical leave support from the Southeastern Universities Research Association (JLG).

\newpage

\nopagebreak \squeezetable \begin{table}
\begin{tabular}{cclc} Multiplet & Example Decay & \multicolumn{1}{c}{$V_\mu$}
& Suppression \\ [5pt]\hline \\ [5pt] $(0^-,1^-)$ & $1^-\to 0^-$ &
$\xi(v_a\cdot v_b)\epsilon_{\mu\nu\alpha\beta}\epsilon^\nu v_a^\alpha
v_b^\beta$ &  $\frac{1}{m_c}$ \\ [5pt] $(0^+,1^+)$ & $1^+\to 0^-$&
$\tau_{1/2^+}(v_a\cdot v_b)\left[\left(v_a\cdot v_b-1\right) \epsilon_\mu-
\epsilon\cdot v_b v_{a\mu}\right]$ &  $\frac{1}{m_c^2}$ \\ [5pt] $(1^+,2^+)$ &
$2^+\to 0^-$ & $\tau_{3/2^+}(v_a\cdot
v_b)\epsilon^{\rho\nu}v_{b\nu}\epsilon_{\mu\rho\alpha\beta}v_a^\alpha
v_b^\beta$ &  $\frac{1}{m_c^2}$ \\ [5pt] $(1^-,2^-)$ & $2^-\to 0^-$ &
$\tau_{3/2^-}(v_a\cdot v_b)\epsilon^{\rho\nu} \left[\left(v_a\cdot
v_b-1\right)g_{\rho\mu}v_{b^\nu}- v_{b\rho}v_{b\nu}v_{a\mu}\right]$ & 
$\frac{1}{m_c^3}$ \\ [5pt] $(0^-,1^-)^\prime$ & $1^{-\prime}\to 0^-$ &
$\tau_{1/2^-}(v_a\cdot v_b) \left(v_a\cdot v_b-1\right)   
\epsilon_{\mu\nu\alpha\beta}\epsilon^\nu v_a^\alpha v_b^\beta$ & 
$\frac{1}{m_c^3}$ \\ [5pt] \end{tabular} 
\vspace*{3mm}
\caption{Matrix elements of the heavy
quark vector current for a  few selected decays, and the $1/m_c$ suppression
factors associated with them. The polarization vector (tensor) of  the initial
state meson is denoted by $\epsilon$.\label{suppressions}} \end{table}
\nopagebreak

\squeezetable
\begin{table}
\begin{tabular}{lclllcrrr}\\
Decay & $k$ &
\multicolumn{3}{c}{$\bar{c} d$ states}& $k$ & \multicolumn{3}{c}{$\bar{b} d$ states} \\ 
\cline{3-5}\cline{7-9}\\
   & (MeV) & $\Gamma(\kappa^q=0)$  & $\Gamma(\kappa^q=0.45)$ & 
$\Gamma(\kappa^q=0.55)$ &(MeV)  
   & $\Gamma(\kappa^q=0)$  & $\Gamma(\kappa^q=0.45)$ & 
$\Gamma(\kappa^q=0.55)$ \\   \hline \\ [5pt]
 $1^- \to 0^-$ & 136
  &   50 $\pm $  2 eV 
  &  904 $^{+ 25}_{- 24}$   eV 
  &    1.5 $\pm $  0.0 keV &  45
  &   37 $^{+  5}_{-  4}$   eV 
  &  182 $^{+ 22}_{- 21}$   eV 
  &  228 $^{+ 28}_{- 26}$   eV 
 \\[5pt]
 $1^+ \to 0^+$ & 
127 
  &  510 $^{+ 257}_{- 194}$   eV 
  &    1.6 $^{+   0.8}_{-   0.6}$   keV
  &    1.9 $^{+   0.9}_{-   0.7}$   keV & 
 40
  &    4.5 $^{+  10.7}_{-   4.0}$   eV 
  &    1.0 $^{+   2.3}_{-   0.9}$   eV 
  &    2.9 $^{+   6.7}_{-   2.6}$   eV 
 \\[5pt]
 $2^+ \to 1^+$ & 
 32 
  &    7.0 $^{+  22.7}_{-   6.7}$   eV 
  &    0.6 $^{+   1.8}_{-   0.5}$   eV 
  &    0.1 $^{+   0.4}_{-   0.1}$   eV &
 20
  &    1.9 $^{+  13.3}_{-   1.9}$   eV 
  &    5.5 $^{+  37.9}_{-   5.5}$   eV 
  &    6.5 $^{+  45.1}_{-   6.5}$   eV 
 \\[5pt]
 $2^- \to 1^-$ & 
 30
  &   11 $^{+  38}_{-  10}$   eV 
  &   20 $^{+  71}_{-  19}$   eV 
  &   23 $^{+  80}_{-  22}$   eV &
  0
  &    0.0 $^{+   0.1}_{-   0.0}$   eV 
  &    0.0 $\pm $   0.0 eV 
  &    0.0 $\pm $   0.0 eV 
 \\[5pt]
 $1^{-'} \to 0^{-'}$ & 
117
  &  153 $^{+  91}_{-  65}$   eV 
  &  311 $^{+ 149}_{- 119}$   eV 
  &  590 $^{+ 290}_{- 229}$   eV  & 
 40
  &   15 $^{+  34}_{-  13}$   eV 
  &  101 $^{+ 234}_{-  88}$   eV 
  &  131 $^{+ 303}_{- 114}$   eV 
 \\ [5pt] \hline \\
 & &\multicolumn{3}{c}{$\bar{c} u$ states}& &
\multicolumn{3}{c}{$\bar{b} u$
 states} \\
\cline{3-5}\cline{7-9}\\[5pt]
 $1^- \to 0^-$ & 
137
  &    7.3 $\pm $  0.2 keV
  &   26 $\pm $  1 keV
  &   32 $\pm $  1 keV & 45
  &   84 $\pm 10$   eV 
  &  572 $^{+ 71}_{- 65}$   eV 
  &  740 $^{+ 92}_{- 85}$   eV
 \\[5pt]
 $1^+ \to 0^+$ & 
127
  &    1.9 $^{+   1.0}_{-   0.7}$   keV
  &  102 $^{+  62}_{-  43}$   eV 
  &    6.6 $^{+   6.6}_{-   3.5}$   eV  & 
 40
  &    3.0 $^{+   7.0}_{-   2.6}$   eV 
  &   21 $^{+  48}_{-  18}$   eV 
  &   36 $^{+  83}_{-  31}$   eV 
 \\[5pt]
 $2^+ \to 1^+$ & 
 32
  &   77 $^{+ 245}_{-  73}$   eV 
  &  158 $^{+ 505}_{- 150}$   eV 
  &  180 $^{+ 576}_{- 171}$   eV &
 20
  &    2.8 $^{+  19.2}_{-   2.8}$   eV 
  &   12.8 $^{+  88.5}_{-  12.8}$   eV 
  &   16.0 $^{+ 110.8}_{-  16.0}$   eV 
 \\[5pt]
 $2^- \to 1^-$ & 
 30
  &   40 $^{+ 141}_{-  38}$   eV 
  &   15 $^{+  52}_{-  14}$   eV 
  &   11 $^{+  38}_{-  10}$   eV &
  0
  &    0.0 $\pm $   0.1 eV 
  &    0.0 $\pm $   0.0 eV 
  &    0.0 $^{+   0.1}_{-   0.0}$   eV 
 \\[5pt]
 $1^{-'} \to 0^{-'}$ & 
117
  &    3.1 $^{+   1.7}_{-   1.2}$   keV
  &   13.3 $^{+   7.1}_{-   5.3}$   keV
  &   16.5 $^{+   8.8}_{-   6.6}$   keV & 
 40
  &   26 $^{+  60}_{-  23}$   eV 
  &  308 $^{+ 713}_{- 269}$   eV 
  &  412 $^{+ 955}_{- 360}$   eV 
 \\[5pt] 
\end{tabular}\vspace*{3mm}
\caption{Radiative decay widths of non-strange heavy mesons: 
only transitions within heavy quark spin multiplets are shown. 
\label{nonstrange}}
\end{table}

\squeezetable
\begin{table}
\begin{tabular}{lrrrrrrrr}\\
Quantity & \multicolumn{1}{c}{MS} & \multicolumn{1}{c}{Eichten} & 
\multicolumn{1}{c}{Pham} & 
\multicolumn{1}{c}{Rosner} & 
\multicolumn{1}{c}{Kamal} & \multicolumn{1}{c}{Cheng} & \multicolumn{2}{c}{this 
work} \\
\cline{8-9}
 & & & & & & & $\kappa^q=0.45$ & $\kappa^q=0.55$ \\\hline
BR$(D^{*+}\to D^+\pi^0)$ & 31.2 & 28.5 & 29.4 & 30.9 & 30.0 &
31.2 & 30.5 & 30.3  \\[5pt]
BR$(D^{*+}\to D^0\pi^+)$ & 67.5 & 68.5 & 64.7 & 67.8 & 68.0 &
67.3 & 68.5 & 68.1  \\[5pt]
BR$(D^{*+}\to D^+\gamma)$ & 1.3 & 3.0 & 5.9 & 1.3 & 2.0 & 1.5 &
1.0 & 1.5  \\[5pt]
$\Gamma(D^{*+}\to {\rm all})$ (keV) & 79.0 & 78.0 & 142.8 & 83.9 & 86.4 &
141.0  & 94.3 & 94.9 \\ [5pt]
BR$(D^{*0}\to D^0\pi^0)$ & 64.3 & 55.2 & 71.4 & 70.6 & 66.0 &
66.7 & 61.5 & 56.5  \\ [5pt]
BR$(D^{*0}\to D^0\gamma)$ & 35.7 & 44.8 & 28.6 & 29.4 & 34.0 &
33.3 & 38.5 &  43.5 \\[5pt]
$\Gamma(D^{*0}\to {\rm all})$ (keV) & 59.4 & 78.6 & 120.4 & 56.2 & 64.2 &
102.0  & 67.6 & 73.6 \\ [5pt]
$\Gamma(D_s^{*+}\to D_s^+\gamma)$ (keV) & & & & & 0.21 & 0.3 &
0.2 & 0.3 \\ \end{tabular}
\vspace*{3mm}
\caption{Total widths and branching
  ratios for charmed vector mesons. The results shown are from Miller and
Singer (MS) \protect\cite{miller}; Eichten {\it et al.} (Eichten) 
\protect\cite{eichten}; Pham
\protect\cite{pham}; Rosner \protect\cite{rosner}; Kamal and Xu (Kamal) 
\protect\cite{kamal}; Cheng {\it et al.}
(Cheng) \protect\cite{cheng}; and the present work. The numbers in the last two 
columns correspond to $g_A^q=0.8$, and
$\kappa^q=0.45$, $\kappa^q=0.55$, respectively, and are calculated using the 
appropriate `central' values shown in tables 
\ref{nonstrange} and
\ref{strange}. \label{otherresults}}
\end{table}
\newpage

\squeezetable
\begin{table}
\begin{tabular}{lclllcrrr}\\
Decay & $k$ &
\multicolumn{3}{c}{$\bar{c} s$ states}& $k$ & \multicolumn{3}{c}{$\bar{b}
s$ states} \\  \cline{3-5}\cline{7-9}\\
   & (MeV) & $\Gamma(\kappa^q=0)$  & $\Gamma(\kappa^q=0.45)$ & 
$\Gamma(\kappa^q=0.55)$ &(MeV)  
   & $\Gamma(\kappa^q=0)$  & $\Gamma(\kappa^q=0.45)$ & 
$\Gamma(\kappa^q=0.55)$ \\   \hline \\[5pt]
 $1^- \to 0^-$ &
139
  &  101 $\pm $   3 eV 
  &  165 $\pm $   4 eV 
  &  321 $^{+   9}_{-   8}$   eV  &
 47
  &   35 $^\pm 3$   eV 
  &  113 $^\pm 10$   eV 
  &  136 $^\pm 12$   eV 
 \\ [5pt]
 $1^+ \to 0^+$ &
127
  &  547 $^{+ 276}_{- 209}$   eV 
  &    1.1 $^{+   0.5}_{-   0.4}$   keV
  &    1.2 $^{+   0.6}_{-   0.5}$   keV &
 40
  &    4.0 $^{+   9.4}_{-   3.5}$   eV 
  &    0.0 $\pm $   0.1 eV 
  &    0.0 $\pm $   0.1 eV 
 \\[5pt]
 $2^+ \to 1^+$ &
 38
  &   13 $^{+  33}_{-  12}$   eV 
  &    4.8 $^{+  12.2}_{-   4.3}$   eV 
  &    3.5 $^{+   8.9}_{-   3.1}$   eV &
 30
  &    5.7 $^{+  20.3}_{-   5.5}$   eV 
  &   11.7 $^{+  41.7}_{-  11.2}$   eV 
  &   13.3 $^{+  47.5}_{-  12.8}$   eV 
 \\[5pt]
 $2^- \to 1^-$ &
 40
  &   26.1 $^{+  60.5}_{-  22.8}$   eV 
  &   38.5 $^{+  89.1}_{-  33.6}$   eV 
  &   41.6 $^{+  96.2}_{-  36.3}$   eV &
  0
  &    0.0 $^{+   0.8}_{-   0.0}$   eV 
  &    0.0 $^{+   0.2}_{-   0.0}$   eV 
  &    0.0 $^{+   0.2}_{-   0.0}$   eV 
 \\[5pt]
 $1^{-\prime} \to 0^{-\prime}$ &
117
  &  177 $^{+ 104}_{-  74}$   eV 
  &   21 $^{+   8}_{-   7}$   eV 
  &   72 $^{+  32}_{-  28}$   eV  &
 40
  &   13 $^{+  31}_{-  12}$   eV 
  &   53 $^{+ 124}_{-  47}$   eV 
  &   66 $^{+ 153}_{-  58}$   eV 
\\[5pt] \hline \\
 $0^+ \to 1^-$ &
253
  &   24.9 $\pm $   1.9 keV
  &   14.5 $\pm $   0.9 keV
  &   16.2 $\pm $   1.0 keV & 324
  &   35.9 $\pm $   2.6 keV
  &   20.1 $\pm $   1.4 keV
  &   22.6 $\pm $   1.6 keV \\[5pt]
 $1^+ \to 0^-$ &
483
  &   17.2 $\pm $   0.7 keV
  &   10.3 $\pm $   0.6 keV
  &   11.4 $\pm $   0.6 keV & 406
  &   15.9 $\pm $   1.0 keV
  &    9.0 $\pm $   0.6 keV
  &   10.1 $\pm $   0.7 keV
 \\[5pt]
 $1^+ \to 1^-$ &
366
  &   25.1 $\pm $   1.4 keV
  &   14.0 $\pm $   0.8 keV
  &   15.8 $\pm $   0.9 keV & 362
  &   27.5 $\pm $   1.8 keV
  &   15.4 $\pm $   1.1 keV
  &   17.3 $\pm $   1.2 keV
 \\ [5pt]\hline \\
 $1^+ \to 0^-$ &
503
  &   25.2 $\pm $   0.5 keV
  &   31.1 $\pm $   0.8 keV
  &   30.0 $\pm $   0.7 keV & 406
  &   30.5 $^{+   0.1}_{-   0.2}$   keV
  &   38.3 $^{+   0.0}_{-   0.2}$   keV
  &   36.8 $^{+   0.0}_{-   0.2}$   keV
 \\[5pt]
 $1^+ \to 1^-$ &
387
  &   14.6 $\pm $   0.2 keV
  &   22.8 $\pm $   1.2 keV
  &   21.0 $\pm $   1.0 keV & 362
  &   15.8 $\pm $   0.4 keV
  &   23.4 $\pm $   1.5 keV
  &   21.8 $\pm $   1.2 keV
 \\[5pt]
 $2^+ \to 0^-$ &
534
  &    1.6 $\pm $   0.2 keV
  &    8.6 $\pm $   1.0 keV
  &    6.9 $\pm $   0.8 keV & 434
  &  795 $^{+ 158}_{- 139}$   eV 
  &    4.2 $^{+   0.8}_{-   0.7}$   keV
  &    3.4 $\pm $   0.7 keV
 \\[5pt]
 $2^+ \to 1^-$ &
419
  &   41.5 $\pm $   0.0 keV
  &   55.9 $^{+   0.5}_{-   0.6}$   keV
  &   53.0 $^{+   0.4}_{-   0.5}$   keV & 390
  &   46.2 $^{+   0.4}_{-   0.6}$   keV
  &   61.1 $^{+   1.2}_{-   1.4}$   keV
  &   58.2 $^{+   1.0}_{-   1.2}$   keV
 \\ \hline \\ 
 $1^+ \to 0^+$ &
150
  &  103 $^{+  39.7}_{-  32.4}$   eV 
  &    1.4 $\pm $   0.5 keV
  &    1.0 $\pm $   0.4 keV &  40
  &    2.2 $^{+   5.0}_{-   1.9}$   eV 
  &   30.4 $^{+  69.9}_{-  26.6}$   eV 
  &   23.0 $^{+  52.9}_{-  20.1}$   eV 
 \\[5pt]
 $1^+ \to 1^+$ &
 25
  &    0.3 $^{+   1.2}_{-   0.3}$   eV 
  &    3.7 $^{+  17.3}_{-   3.7}$   eV 
  &    2.8 $^{+  13.1}_{-   2.8}$   eV &  0
  &    0.0 $^{+   0.1}_{-   0.0}$   eV 
  &    0.0 $^{+   1.9}_{-   0.0}$   eV 
  &    0.0 $^{+   1.5}_{-   0.0}$   eV 
 \\[5pt]
 $2^+ \to 0^+$ &
186
  &    0.1 $\pm $   0.1 eV 
  &    3.8 $^{+   3.1}_{-   1.9}$   eV 
  &    2.7 $^{+   2.2}_{-   1.3}$   eV &  70
  &    0.0 $\pm $   0.0 eV 
  &    0.0 $\pm $   0.0 eV 
  &    0.0 $\pm $   0.0 eV 
 \\[5pt]
 $2^+ \to 1^+$ &
 62
  &   12.0 $^{+  14.7}_{-   8.1}$   eV 
  &  168 $^{+ 203}_{- 113}$   eV 
  &  127 $^{+ 153}_{-  85.1}$   eV &  30
  &    1.4 $^{+   4.9}_{-   1.3}$   eV 
  &   19.5 $^{+  68.7}_{-  18.7}$   eV 
  &   14.7 $^{+  51.9}_{-  14.1}$   eV 
 \\ [5pt]\hline \\
 $0^{-'} \to 1^-$ &
450
  &  157 $^{+  24.7}_{-  22.1}$   eV 
  &    1.4 $\pm $   0.1 keV
  &    1.1 $\pm $   0.1 keV & 492
  &  106 $^{+  33.3}_{-  33.5}$   eV 
  &    1.4 $\pm $   0.1 keV
  &    1.0 $\pm $   0.1 keV
 \\[5pt]
 $1^{-'} \to 0^-$ &
655
  &   59.2 $^{+  29.7}_{-  22.8}$   eV 
  &    0.1 $^{+   8.7}_{-   0.1}$   eV 
  &    1.3 $^{+  13.2}_{-   0.5}$   eV & 571
  &    0.2 $^{+   3.7}_{-   0.2}$   eV 
  &  200 $^{+  62.3}_{-  64.4}$   eV 
  &  137 $^{+  48.0}_{-  51.0}$   eV 
 \\[5pt]
 $1^{-'} \to 1^-$ &
548
  &    9.4 $^{+   11.9}_{-  7.5}$   eV 
  &  482 $^{+  98.6}_{-  96.4}$   eV 
  &  346 $^{+  78.5}_{-  78.2}$   eV & 528
  &   28.2 $^{+  16.3}_{-  20.2}$   eV 
  &  696 $^{+ 123}_{- 112}$   eV 
  &  511 $^{+ 101}_{-  93.3}$   eV 
 \\ [5pt]\hline \\
 $0^{-'} \to 1^+$ &
 98
  &    3.3 $\pm $   0.6 keV
  &    4.0 $\pm $   0.7 keV
  &    3.8 $\pm $   0.7 keV & 138
  &    4.6 $\pm $   0.5 keV
  &    5.6 $\pm $   0.7 keV
  &    5.4 $\pm $   0.7 keV
 \\[5pt]
 $1^{-'} \to 0^+$ &
328
  &    2.4 $\pm $   0.0 keV
  &    2.6 $\pm $   0.1 keV
  &    2.5 $\pm $   0.1 keV & 216
  &    2.1 $\pm $   0.1 keV
  &    2.6 $\pm $   0.1 keV
  &    2.5 $\pm $   0.1 keV
 \\[5pt]
 $1^{-'} \to 1^+$ &
211
  &    4.0 $\pm $   0.2 keV
  &    4.9 $\pm $   0.2 keV
  &    4.8 $\pm $   0.2 keV & 177
  &    3.7 $\pm $   0.3 keV
  &    4.6 $\pm $   0.3 keV
  &    4.5 $\pm $   0.3 keV
 \\ [5pt]\hline \\    
 $0^{-'} \to 1^+$ &
 74
  &    4.6 $\pm $   1.2 keV
  &    4.3 $\pm $   1.1 keV
  &    4.3 $\pm $   1.1 keV & 138
  &    9.0 $\pm $   1.4 keV
  &    8.2 $\pm $   1.3 keV
  &    8.4 $\pm $   1.3 keV
 \\[5pt]
 $0^{-'} \to 2^+$ &
 37
  &    0.0 $^{+   0.2}_{-   0.0}$   eV 
  &    0.1 $^{+   1.1}_{-   0.1}$   eV 
  &    0.1 $^{+   0.9}_{-   0.1}$   eV & 109
  &    5.3 $^{+   6.6}_{-   3.3}$   eV 
  &   33.1 $^{+  41.8}_{-  20.8}$   eV 
  &   26.4 $^{+  33.3}_{-  16.5}$   eV 
 \\[5pt]
 $1^{-'} \to 1^+$ &
188
  &    2.1 $\pm $   0.3 keV
  &    2.1 $\pm $   0.4 keV
  &    2.1 $\pm $   0.3 keV & 177
  &    2.0 $\pm $   0.3 keV
  &    2.0 $\pm $   0.4 keV
  &    2.0 $\pm $   0.3 keV
 \\[5pt]
 $1^{-'} \to 2^+$ &
153
  &    8.1 $\pm $   1.1 keV
  &    7.5 $\pm $   1.1 keV
  &    7.6 $\pm $   1.1 keV & 148
  &    8.1 $\pm $   1.2 keV
  &    7.5 $\pm $   1.2 keV
  &    7.6 $\pm $   1.2 keV
 \\ [5pt]
 \end{tabular}
\vspace*{3mm}
\caption{Radiative decay widths of strange heavy mesons.
\label{strange}}
\end{table}

\end{document}